# Orbital reconstruction and two-dimensional electron gas at the LaAlO$_3$/SrTiO$_3$ interface

Salluzzo, M.[1], Ghiringhelli, G.[2], Cezar, J. C.[3], Brookes, N. B.[3], Bisogni, V.[3], De Luca, G. M.[1], Richter, C.[4], Thiel, S.[4], Mannhart, J.[4], Huijben, M.[5], Brinkman, A.[5], Rijnders, G.[5]

[1] *CNR-INFM COHERENTIA, Complesso MonteSantangelo via Cinthia, 80126 Napoli, Italy*

[2] *CNR-INFM COHERENTIA and Dipartimento di Fisica, Politecnico di Milano, piazza Leonardo da Vinci 32, I-20133 Milano, Italy*

[3] *European Synchrotron Radiation Facility, 6 rue Jules Horowitz, B.P. 220, F-38043 Grenoble Cedex, France*

[4] *Experimental Physics VI, Center for Electronic Correlations and Magnetism, Institute of Physics, University of Augsburg, D-86135 Augsburg, Germany.*

[5] *Faculty of Science & Technology and MESA+ Institute for Nanotechnology, University of Twente, PO Box 217, 7500 AE Enschede, The Netherlands*

*(Dated: December 8, 2008)*

## Abstract

Conventional two-dimensional electron gases are realized by engineering the interfaces between semiconducting compounds. In 2004, Ohtomo and Hwang discovered that an electron gas can be also realized at the interface between large gap insulators made of transition metal oxides [1]. This finding has generated considerable efforts to clarify the underlying microscopic mechanism. Of particular interest is the LaAlO$_3$/SrTiO$_3$ system, because it features especially striking properties. High carrier mobility [1], electric field tuneable superconductivity [2] and magnetic effects [3], have been found. Here we show that an orbital reconstruction is underlying the generation of the electron gas at the LaAlO$_3$/SrTiO$_3$ n-type interface. Our results are based on extensive investigations of the electronic properties and of the orbital structure of the interface using X-ray Absorption Spectroscopy. In particular we find that the degeneracy of the Ti 3d states is fully removed, and that the Ti 3d$_{xy}$ levels become the first available states for conducting electrons.

Interfaces between transition metal oxides can exhibit electronic properties that are absent in the individual layers. A prominent example is the electronic conduction at the $TiO_2$/LaO *n*-type interfaces in $LaAlO_3$/$SrTiO_3$ (LAO/STO) [1]-[3], $LaTiO_3$/$SrTiO_3$ (LTO/STO) [4] and $LaVO_3$/$SrTiO_3$ (LVO/STO) [5] bilayers composed by excellent insulators. These heterostructures are believed to avoid electronic instabilities at their interfaces by an "electronic reconstruction" mechanism that is expected to occur in addition to the classical structural reconstructions. Consequently, the properties of transition metal oxide interfaces are thought to be sensitively controlled by polar discontinuities, band-bending, epitaxial strain and related phenomena [6].

The LVO/STO and LAO/STO bilayers only generate a conducting interface if the thicknesses of the LVO and LAO films reach a critical value, which are 5 [2] and 4 unit cells (uc) [7] respectively. This behaviour is consistent with the "polarization catastrophe" mechanism, which attributes the induction of mobile charges at the interface to the large electrostatic energy of thick LVO or LAO layers. This energy increases with the film thickness [3, 8], and it is composed of a Coulomb energy term and of the elastic energy of the crystal lattices, which is controlled by the finite lattice mismatch. Therefore, modifications of both structural and electronic properties of the $LaAlO_3$ and $SrTiO_3$ layers in the LAO/STO system may be induced at the interface, raising the question, whether the formation of an electron gas at the interface is simply a result of the induction of charge carriers or whether it is associated with further changes of the electronic or lattice structure of the oxide layers.

To resolve this problem, we have applied X-ray Absorption Spectroscopy (XAS) as a probe of the interface electronic states of the LAO/STO system. X-ray absorption spectroscopy is able of identifying the unoccupied density of states of buried interfaces with chemical and orbital sensitivity. It has recently been used with great success to shed light on the interfaces electronic states in manganite/cuprate bilayers [9].

The XAS measurements have been performed at the ID08 beamline of the European Synchrotron Radiation Facility. XAS spectra were collected at the Ti-$L_{2,3}$ absorption edge of *n*-doped interfaces in LAO/STO heterostructures grown by pulsed laser deposition. The spectra were measured as a function of the thickness of the LAO films. Details concerning the method of preparation, structural, transport and morphological properties of these samples can be found in [2]-[3]. In particular, the analyzed samples were grown at the University of Augsburg at an oxygen pressure of $8 \times 10^{-5}$ mbar and at the University of Twente at an oxygen pressure of $2 \times 10^{-3}$ mbar. The former show low temperature superconductivity [2], for the latter ones magnetic effects were reported [3]. The LAO/STO films were oxygenated *in-situ* at an oxygen pressure of ~0.5 bar. By using scanning probe microscopy this procedure has been demonstrated to create a conductive layer only at the interface [10].

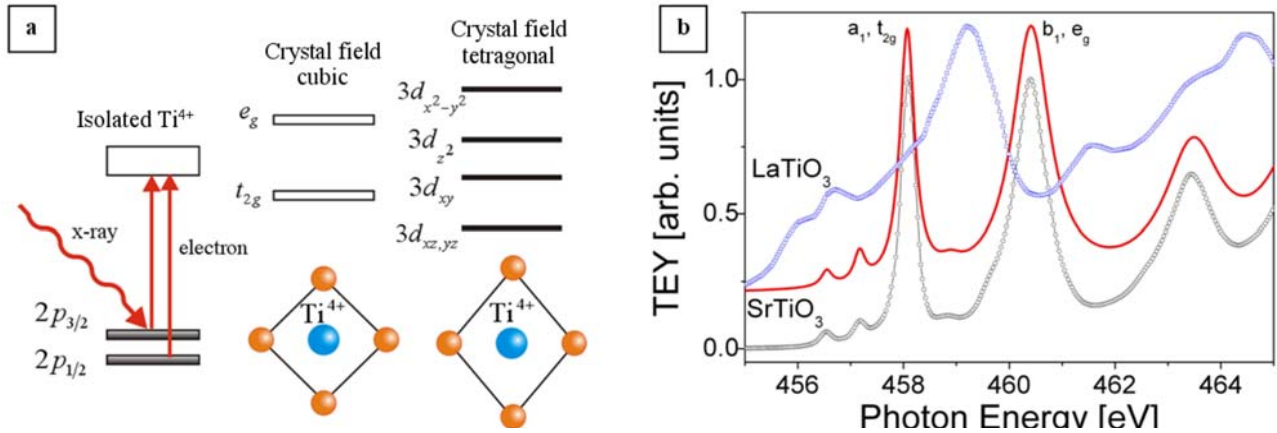

**Figure 1 X-ray Absorption Spectroscopy on STO and LTO crystals:** (a) a sketch of the XAS process for 2p→3d transitions. The available, empty 3d $Ti^{4+}$ states are depicted for cubic (middle) and tetragonal (right) crystal fields. (b) XAS spectra, acquired with E//ab ($I_{ab}$) on a $TiO_2$ terminated $SrTiO_3$ single crystal (black circles). The blue circles show reference XAS spectra taken on a $LaTiO_3$ single crystal. The data present the total electron yield in arbitrary units. The red, continuous line presents the result of the atomic multiplet simulation for STO using the Missing package software [11].

In X-ray absorption, photons excite core electrons to the unoccupied states of the solid. In $SrTiO_3$, the first available unoccupied states are the empty 3d levels of the $Ti^{4+}$ ions [Fig.1a]. All the features of the measured $L_{2,3}$ Ti-edge spectra agree with the results of atomic multiplet calculations, which were performed using the "Missing" scientific package [11] [Fig. 1b]. The first two main spectral peaks, $a_1$ and $b_1$, are attributed to $2p_{3/2}\rightarrow$ 3d transitions and have contribution from $e_g$ ($d_{xz}$, $d_{yz}$, $d_{xy}$) and $t_{2g}$ ($d_{z^2}$, $d_{x^2-y^2}$) levels respectively. Here, the *z*-direction refers to the surface normal. The XAS spectra are strongly sensitive to the valence of the Ti ions [12]. Indeed, data taken on $LaTiO_3$, the Ti-sublattice of which comprises $Ti^{3+}$ only, differ considerably from those taken on STO [Fig. 1b]. Moreover, the XAS process is strongly influenced by the splitting of the 3d levels and therefore provides information on the crystal field. In fact, by using linearly polarized light, it is possible to select an in-plane ($d_{xy}$, $d_{x^2-y^2}$) or an out-of-plane ($d_{xz}$, $d_{yz}$, and $d_{z^2}$) orbital as final state. The linear dichroism, obtained by subtracting XAS spectra with out-of-plane ($I_c$) and in-plane polarization ($I_{ab}$), is particularly sensitive to any distortion of the $TiO_6$ octahedra and therefore provides valuable information on the eventual presence of orbital reconstructions.

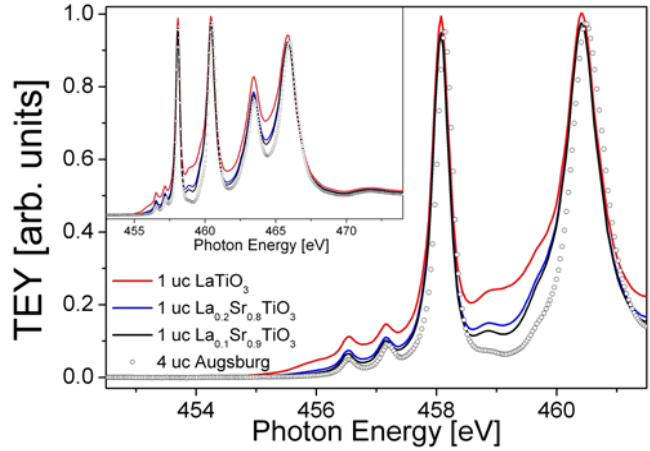

**Figure 2 X-ray absorption spectrum measured at an *n*-type LAO/STO interface in comparison to the calculated spectra of a single $La_xSr_{1-x}TiO_{3-d}$ monolayer:** the experimental results (black circles) are obtained on a 4 uc LAO/STO bilayer, the lines show the calculated spectra of STO covered by one monolayer of $La_xSr_{1-x}TiO_{3-d}$ with *x*=1 (red), *x*=0.2 (blue) and *x*=0.1 (black).

Bulk sensitive fluorescence yield and interface sensitive total electron yield XAS acquisition modes were used simultaneously. Fluorescence Yield (FY) XAS spectra of LAO/STO interfaces, composed of 2 uc, 4 uc, 8 uc and 12 uc LAO layers, of insulating STO crystals and of conducting Nb-doped (1% at.) STO crystals, are very similar. Moreover, as function of the polarization the data do not show any appreciable difference. The Total Electron Yield (TEY) XAS spectra, however, turns out to vary among the samples and depend significantly on the polarization. The differences between FY and TEY data are related to the different sampling depths of the two techniques. For the soft x-ray energies used (<500 eV), the sampling depth in the FY mode is several tens of nm, while in the TEY mode it is between 1.5 and 3.0 nm [13]. Thus, the TEY mode is mainly probing the final STO interface layers. Therefore, the differences of the two data-sets provide clear evidence that the electronic states at the interfaces of the LAO/STO bilayer differ from the bulk states.

These measurements reveal, moreover, that the interface electronic states are not associated with a formation of a $La_xSr_{1-x}TiO_{3-d}$ solid solution. Fig. 2 compares data measured at a conducting LAO/STO sample (4 uc) with calculated spectra of a hypothetical complete interface layer of $La_xSr_{1-x}TiO_{3-d}$. The latter are obtained by appropriately combining the measured TEY of $LaTiO_3$ and $SrTiO_3$ crystals. Fig. 2 reveals that a complete $LaTiO_3$ or $La_xSr_{1-x}TiO_{3-d}$ layers ($x>0.1$) are unable to match the experimental results, in agreement with previous electron energy loss spectroscopy measurements [2].

TEY XAS spectra of LAO/STO heterostructures measured as function of the number of LAO layers provide key information on the electronic properties of the interface. First, with increasing LAO thickness, the main peaks, in particular the peak $b_1$ of the $I_{ab}$ and $I_c$ spectra, shift to higher energies and their position saturates above 4 uc [Fig. 3a]. A second key result comes from the measurements of the linear dichroism, i.e. the differences among the TEY XAS signals taken with the two polarizations [Fig. 3b]. A marked change in the linear dichroism is found as function of the LAO thickness. This result has to be associated to a change of the anisotropy in the 3d electronic states and in particular to a variation of the splitting between the in-plane and out-of-plane $t_{2g}$ and $e_g$ states. In a perfectly symmetric $Ti^{4+}$ system, as in cubic STO with undistorted $TiO_6$ octahedra, the absolute dichroism needs to disappear. Dichroism can be generated by a deviation from the cubic symmetry, for example by a tetragonal distortion, which causes an energy splitting of the $t_{2g}$ and $e_g$ states. The energy splitting can be obtained by treating it as a fit-parameter in fitting the calculated spectra to the $I_{ab}$ and $I_c$ XAS data and to the corresponding dichroism. Remarkably, STO and metallic Nb-doped STO reference samples show a dichroism opposite to the one of the LAO/STO interface [Fig. 3b]. In the case of STO and Nb-doped STO, the experimental results are well reproduced by a positive energy splitting of about 20 meV between the $d_{xy}$ and the ($d_{xz}$, $d_{yz}$) orbitals, and of 40 meV between the $d_{x^2-y^2}$ and $d_{z^2}$ states. These values correspond to an elongation of the $TiO_6$ octahedra along the surface normal [14]. After depositing a few unit cells of $LaAlO_3$ on such a surface, however, the dichroism changes sign [Fig. 3b]. Now the in-plane $d_{xy}$, and $d_{x^2-y^2}$ orbitals have a smaller energy than the out-of-plane ($d_{xz}$, $d_{yz}$) and $d_{z^2}$ orbitals. The energy differences amount to ~50 meV and 100 meV, respectively. The splitting of the energy levels, and the shift of the $a_1$ and $b_1$ peaks, occur already in the insulating 2 uc sample, but become much more pronounced if the LAO thickness exceeds the critical value of $d \geq 4$ uc for the interface to become conducting.

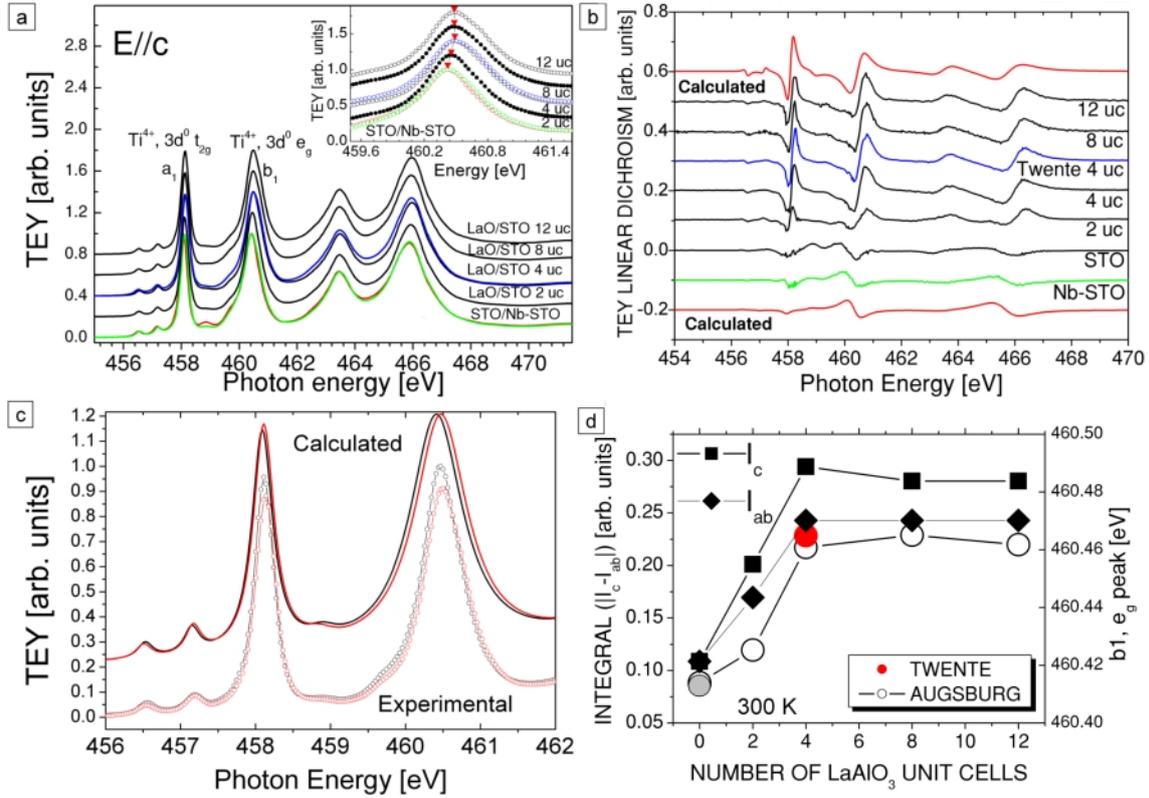

**Figure 3 TEY X-ray Absorption Spectroscopy on LAO/STO interface:** (a) TEY Ti $L_{23}$ edge spectra with E//c polarization and (b) linear dichroism acquired at 300 K on STO (red line), Nb-STO (green line) and LAO/STO as function of the LAO thickness (black lines, Augsburg samples grown at $8 \times 10^{-5}$ mbar, blue lines Twente 4 uc sample grown at $2 \times 10^{-3}$ mbar). The inset of (a) shows that with increasing LAO thickness the peak $b_1$ shifts to higher energies. In (b) the calculated dependence of the linear dichroisms reproducing the data on conducting LAO/STO (upper curve) and on bare STO (lower curve) are shown as red lines. (c) Comparison between $I_{ab}$ (black) and $I_c$ (red) experimental TEY spectra on 4 uc LAO/STO (open circles) and calculated ones for a bilayer (continuos lines). (d) the integral of the absolute dichroism (open circles: Augsburg samples deposited at $8 \times 10^{-5}$ mbar; filled red circle: Twente sample grown at $2 \times 10^{-3}$ mbar; filled grey circle is the data on Nb-doped STO) and the position of the $b_1$ peaks from the $I_{ab}$ (filled diamonds) and $I_c$ (filled squares) spectra as function of the number of $LaAlO_3$ layers.

The details of the XAS spectra are also well reproduced by using the parameters obtained from the splitting of the levels as determined from the dichroism data. It is noted that not only the energy shift between $I_c$ and $I_{ab}$ spectra, but also the shape of the XAS signal in the region between the $a_1$ and the $b_1$ peaks can be accounted for [Fig. 3c]. Finally, the integral of the absolute dichroism, $\int_{L_{2,3}} |I_c - I_{ab}| dE$, and the position of the $b_1$ peaks from the $I_{ab}$ and $I_c$ spectra also show a sudden change as function of the number of LaAlO$_3$ layers, in particular between 2 and 4 uc, *i.e.* when the interface becomes conducting [Fig. 3d].

We conclude that for LAO thicknesses close to the conductance threshold a pronounced anisotropy of the 3d energy levels is induced. This behaviour is correlated with the formation of the 2D electron gas and therefore reveals that an orbital reconstruction occurs when mobile carriers appear at the interface. The orbital reconstruction is robust and, between 300 K and 9 K, almost temperature independent. Moreover, the reconstruction is extremely similar in samples grown in different oxygen conditions. These results are consistent with signatures of electronic reconstructions that have recently been reported for LAO/STO superlattices investigated by resonant x-ray scattering [15].

It is remarkable that the splitting of the 3d levels at the LAO/STO interface is opposite to the case of TiO$_2$ terminated STO and Nb-STO. If a pure Jahn-Teller distortion was assumed, the data would indicate that at the LAO/STO interface the TiO$_6$ octahedra are compressed along the interface normal. This behaviour would, however, be conflicting with structural data, which suggest an out-of-plane relaxation of STO close to the interface [16]-[18]. Such a relaxation is expected since the La$^{3+}$ ion produce a large, repulsive Coulomb field on the Ti ion, pushing the latter toward the STO bulk, while the oxygen in the octahedra are moved toward the LAO. Our data, therefore, suggest that the orbital distortion cannot be a pure Jahn-Teller one. Describing the generation of the electron gas at the LaTiO$_3$/SrTiO$_3$ interface, Okamoto *et al.* have proposed a ferroelectric-like distortion of the TiO$_6$ [19]. Their calculation also shows that this kind of structural relaxation could provide an orbital ferro-distorsive ordering of the 3d$_{xy}$ Ti states, becoming the lowest empty levels for conducting electrons. Similar conclusion are reported in [20] for the same system and in [21] for the LaAlO$_3$/SrTiO$_3$ superlattice. On the other hand, recent LDA and LDA+U [22] calculations on the LaAlO$_3$/SrTiO$_3$ predict an antiferro-distorsive ordering as found in GdFeO$_3$. In this case the conducting electrons would have only partially a 3d$_{xy}$ character, which would be conflicting with the experimental data that reveal the 3d$_{xy}$ orbital to be the first available state.

The calculations performed on superlattices predict an electronic reconstruction and a conducting interface also for one unit cell LAO thin layer, a result experimentally verified in LAO/STO multilayers [23]. However, the situation of a bilayer differs considerably due to the proximity of the interface to the LaAlO$_3$ surface. Thus the dipole field inside the LaAlO$_3$ can be partially accommodated by relaxation and consequent polar distortion of the lanthanum and oxygen ions [24]. Consequently, until a critical thickness threshold, the polar distortion in the LaAlO$_3$ can compensate partially the electrostatic energy. Above the threshold this mechanism is no more efficient and an orbital reconstruction takes place. Our data demonstrate that an orbital reconstruction indeed occurs in the bilayers only above the threshold. In particular, interface side-bands with 3d$_{xy}$ character appear, to which carriers are transferred to lower the electrostatic energy. This is consistent with the ferro-distorsive ordering of the TiO$_6$ orbital predicted for LaTiO$_3$/SrTiO$_3$ and LaAlO$_3$/SrTiO$_3$ superlattice [19]-[21].

In conclusion, X-ray absorption spectra of LAO/STO bilayers reveal that the generation of the 2D electron gas at the interface between LaAlO$_3$ and SrTiO$_3$ is not related to an oxygen-defective or cation-substituted (La/Sr) STO surface layer. The data show that the formation of the conducting electron gas is associated with an orbital reconstruction occurring at the interface, which causes a splitting of the 3d Ti states and the lowering of the $3d_{xy}$ levels so that these become the first empty Ti states for conducting electrons.

**Acknowledgements**

The authors acknowledge support from the EU under the project Nanoxide, contract n. 033191. J. M., S. T., and C. R. gratefully acknowledge support by the German Science Foundation (SFB 484). The authors are grateful to Jean-Marc Triscone, Ralph Claessen, Hans Hilgenkamp, Dave A. Blank, Ruggero Vaglio and to Antonio Barone for useful discussions.

**METHODS**

The XAS measurements have been performed at the ID08 ESRF beamline in a special cryostat where the sample surface is placed in the vertical plane and is kept in ultra high vacuum condition during the experiment. The linearly polarized x-rays were impinged onto the surface at an incident angle of $\theta=70°$ from the surface normal. Using the vertical (V) polarization of the light, dipole selection rules allows transition to final states in the plane of the sample while, in the horizontal (H) polarization, the electric field of the radiation is mainly parallel to the surface normal. A further correction is applied to the spectra having H polarization, in order to get only the E//c component, $I_c$, using the well known formula $I_H(\theta)=I_{ab}\cos^2(\theta)+I_c\sin^2(\theta)$. $I_H(\theta)$ is the spectra taken at the angle theta with H polarization and $I_{ab}$ is the spectra taken with the same geometry using V polarization. No background subtraction is needed using this procedure. We have acquired for each sample and temperature consecutive spectra with H, V, V and H polarization and averaged them to eliminate any systematic errors.

The simulated XAS spectra have been calculated using the MISSING package [11], based on the Cowan's code (www.esrf.eu), for a $Ti^{4+}$ ion ($3d^0$) in $O_h$ and $D_{4h}$ point symmetry. The optimized parameters were: $10D_q=2.19$ eV, Slater integrals rescaled to 61% of their Hartree-Fock values,

Lorentzian of final states variable from 120 meV to 850 meV (HWHM) for increasing final state energy.

The spectra of $La_xSr_{1-x}TiO_{3-d}$ (1uc)/$SrTiO_3$ of Fig. 2 are determined by combining the experimental $SrTiO_3$ and $LaTiO_3$ data. In particular we used the following formula, valid for the TEY of a bilayer, which takes into account the sampling depth, $d=3nm$:

$$I = xI_\infty^{LTO}\left(1-e^{-1/d(uc)}\right) + (1-x)e^{-1/d(uc)}I_\infty^{STO}$$

where $I_\infty^{STO}$ is the bulk spectra of STO, and $I_\infty^{LTO}$ is the bulk spectra of LTO.